\documentclass[twocolumn,showpacs,preprintnumbers]{revtex4}
\usepackage{graphicx}
\usepackage{dcolumn}
\usepackage{bm}
\usepackage{epsfig}
\usepackage{amsmath,amssymb}

\draft

\begin{document}

\title{Branching ratios of $\alpha $-decay to excited states of even-even
nuclei }
\author{Y. Z. Wang$^1$}
\author{H. F. Zhang$^1$}
\email{zhanghongfei@lzu.edu.cn}
\author{J. M. Dong$^1$}
\author{G. Royer$^2$}
\affiliation{$^1$School of Nuclear Science and Technology, Lanzhou University Lanzhou
730000, People's Republic of China\\
$^2$Laboratoire Subatech, UMR: IN2P3/CNRS-Universit\'e-Ecole des
Mines, Nantes 44, France}
\date{\today }

\begin{abstract}
Branching ratios of $\alpha $-decay to members of the ground state
rotational band and excited 0$^{+}$ states of even-even nuclei are
calculated in the framework of the generalized liquid drop model
(GLDM) by taking into account the angular momentum of the $\alpha
$-particle and the excitation probability of the daughter nucleus.
The calculation covers isotopic chains from Hg to Fm in the mass
regions $180<A<202$ and A$\geq 224$. The calculated branching
ratios of the $\alpha $-transitions are in good agreement with the
experimental data and some useful predictions are provided for
future experiments.
\end{abstract}

\pacs{23.60.+e, 21.10.-k, 21.60.-n}
\maketitle




\section{Introduction}\label{sec1}

The $\alpha $ decay process was first explained by Gamow~\cite%
{Gam28} and by Condon and Gurney~\cite{Con28} in 1928 as a quantum
tunneling effect and it is one of the first examples proving the need to use
the quantum mechanics to describe nuclear phenomena and its correctness. On
the basis of the Gamow's theory, the experimental $\alpha $-decay half-lives
of nuclei can be well explained by both phenomenological and microscopic
models~\cite%
{Del02,lov98,Hod03,Buck92,Buck93,Var94,Wau93,GLDM,RG1,Zh1,ZHANG}. Such $%
\alpha $-decay calculations are mainly concentrated on the favored cases,
e.g. the ground state $\alpha $-transitions of even-even nuclei ($\Delta l=0$%
)~\cite{mol97,Fir96}. Besides the favored $\alpha $-transitions, the ground
state of the parent nucleus can also decay to the excited states of the
daughter nucleus ($\Delta l\neq 0$)~\cite{Fir96}. Recently, there is
increasing interest in two kinds of $\alpha $ transitions of even-even
nuclei from both experimental and theoretical sides, i.e. the $\alpha $%
-decay to excited 0$^{+}$ states and to members of ground state
rotational
band~\cite{Wau94,And00,Ric96,den95,Sob01,Mun01,Kar06,Asa06}. These
$\alpha $ transitions belong to the unfavored case, which are
strongly hindered as
compared with the ground state ones. Theoretically, the hindered $\alpha $%
-transition is an effective tool to study the properties of $\alpha $%
-emitters because it is closely related to the internal structure of nuclei~%
\cite{Wau94,And00,Ric96,den95,Sob01,Mun01,Kar06,Asa06}. However, it is 
difficult to describe quantitatively the unfavored $\alpha $-transitions due
to the influence of both non-zero angular momentum and excitation of
nucleons, especially for $\alpha $-emitters in the neighbourhood of the shell
closures. Although the favored $\alpha $-decay model can be
straightforwardly applied to the unfavored $\alpha $-transition, the
calculated branching ratios usually deviate significantly from the
experimental data. Thus it is necessary to improve the favored $\alpha $%
-decay model to describe the unfavored hindered $\alpha $%
-decay. Experimentally it is also very helpful to make theoretical
predictions on unfavored hindered $\alpha $-transitions for future studies.

The generalized liquid drop model (GLDM) has been successfully used to calculate the half-lives of
the favored $\alpha $-decays from the nuclear ground states of even-even nuclei~%
\cite{GLDM,RG1,Zh1,ZHANG}. As far as we know, the unfavored hindered $\alpha $-transitions 
have not been investigated within the GLDM. In
this paper, the GLDM has been improved by taking into account the influence of
the angular momentum of the $\alpha $-particle and the excitation probability of
the daughter nucleus, investigating the hindered $\alpha $-transitions of
even-even nuclei with mass numbers 180 $<$ A $<$ 202 and A $\geq 224$. The calculated branching ratios of $\alpha $-decays 
are compared with the experimental data and the good agreement allows to provide predictions 
for future experiments. This paper is organized as follows. In section 2 the
theoretical framework is introduced. The numerical results and corresponding
discussions are given in Section 3. In the last section, some conclusions
are drawn.

\section{Methods}\label{sec2}
In the favored $\alpha $-decays, such as the ground state to ground state $%
\alpha $-transitions of even-even nuclei, the angular momentum \textit{l}
carried by the $\alpha $-particle is 0. Thus in the former framework of the
GLDM, the centrifugal potential energy \textit{V}$_{\text{cen}}(r)$ is not
included~\cite{GLDM,RG1,Zh1,ZHANG}. In the case of the ground state
of a parent nucleus to the ground state of its deformed daughter nucleus and
to the excited state \textit{I}$^{+}$, the angular momentum \textit{l}
carried by the $\alpha $-particle is not 0. Thus the centrifugal potential
energy \textit{V}$_{\text{cen}}(r)$ which can no more be neglected has been introduced into the GLDM as
\begin{equation}
V_{\text{cen}}(r)=\frac{\hbar ^{2}}{2\mu }\frac{l(l+1)}{r^{2}}
\end{equation}
where $r$ and $\mu $ are the distance between the two fragments and the reduced mass of the $\alpha$%
-daughter system, respectively.

The macroscopic GLDM energy becomes
\begin{equation}  \label{etot}
E=E_{V}+E_{S}+E_{C}+E_{\text{Prox}}+V_{\text{cen}}(r).
\end{equation}
When the nuclei are separated
\begin{equation}  \label{ev}
E_{V}=-15.494\left \lbrack (1-1.8I_1^2)A_1+(1-1.8I_2^2)A_2\right \rbrack \
\text{MeV},
\end{equation}
\begin{equation}  \label{es}
E_{S}=17.9439\left \lbrack(1-2.6I_1^2)A_1^{2/3}+(1-2.6I_2^2)A_2^{2/3} \right
\rbrack \ \text{MeV},
\end{equation}
\begin{equation}  \label{ec}
E_{C}=0.6e^2Z_1^2/R_1+0.6e^2Z_2^2/R_2+e^2Z_1Z_2/r,
\end{equation}
where $A_i$, $Z_i$, $R_i$ and $I_i$ are the mass numbers, charge numbers,
radii and relative neutron excesses of the two nuclei. $r$ is the distance
between the mass centers. The radii $R_{i}$ are given by
\begin{equation}  \label{radii}
R_i=(1.28A_i^{1/3}-0.76+0.8A_i^{-1/3}) \ \text{fm}.
\end{equation}

For one-body shapes, the volume, surface and Coulomb energies are defined as
\begin{equation}  \label{esone}
E_{V}=-15.494(1-1.8I^2)A \ \text{MeV},
\end{equation}
\begin{equation}  \label{esone}
E_{S}=17.9439(1-2.6I^2)A^{2/3}(S/4\pi R_0^2) \ \text{MeV},
\end{equation}
\begin{equation}  \label{econe}
E_{C}=0.6e^2(Z^2/R_0) \times 0.5\int (V(\theta)/V_0)(R(\theta)/R_0)^3 \sin
\theta d \theta.
\end{equation}
$S$ is the surface of the one-body deformed nucleus. $V(\theta )$ is the
electrostatic potential at the surface and $V_0$ the surface potential of
the sphere.

The surface energy results from the effects of the surface tension forces in
a half space. When there are nucleons in regard in a neck or a gap between
separated fragments an additional term called proximity energy must be added
to take into account the effects of the nuclear forces between the close
surfaces. This term is essential to describe smoothly the one-body to
two-body transition and to obtain reasonable fusion barrier heights. It
moves the barrier top to an external position and strongly decreases the
pure Coulomb barrier.
\begin{equation}
E_{\text{Prox}}(r)=2\gamma \int _{h_{\text{min}}} ^{h_{\text{max}}} \Phi
\left \lbrack D(r,h)/b\right \rbrack 2 \pi hdh,
\end{equation}
where $h$ is the distance varying from the neck radius or zero to the height
of the neck border. $D$ is the distance between the surfaces in regard and $%
b=0.99$~fm the surface width. $\Phi$ is the proximity function of Feldmeier.
The surface parameter $\gamma$ is the geometric mean between the surface
parameters of the two nuclei or fragments. The combination of the GLDM and
of a quasi-molecular shape sequence has allowed to reproduce the fusion
barrier heights and radii, the fission and the $\alpha$ and cluster
radioactivity data.

In the unfavored $\alpha $-transitions, the parent nucleus decays to the
excited states of the daughter nucleus. Thus the excitation energy \textit{E}%
$_{I}^{\ast }$ has influence on the penetration probability of $\alpha$%
-particle through the Coulomb barrier. The $\alpha $ transitions are
assumed to occur from the ground state 0$^{+}$ of an even-even
parent nucleus to the rotational band (0$^{+}$, 2$^{+}$, ...,
\textit{I}$^{+}$, ...) of the ground state of a daughter nucleus.
The $\alpha $ decay of 0$^{+}\longrightarrow I^{+}$ transition
requires that the $\alpha $-particle carries an angular momentum of
\textit{l}= \textit{I} to satisfy angular momentum conservation and
parity conservation. The $\alpha$-decay energy from the ground state
of a parent nucleus is related to the excitation energy of the
\textit{I}$^{+}$ states in the daughter nucleus. It is the
subtraction between the decay energy of the ground state and the
excitation energy of the \textit{I}$^{+}$ state
\begin{equation}  \label{decay energy}
Q_{0^{+}\longrightarrow I^{+}}=Q_{0^{+}\longrightarrow 0^{+}}-E_{I}^{\ast}
\end{equation}

Partial half-life of the ground state of a parent nucleus to each state of a
rotational band of its daughter nucleus can be obtained with different
orbital angular momentum \textit{l} and $\alpha $ decay energy \textit{Q}$%
_{0^{+}\longrightarrow I^{+}}$.

The half-life of a parent nucleus decaying via $\alpha$-emission is
calculated using the WKB barrier penetration probability. The barrier
penetrability \textit{P}$(Q_{\alpha },E_{I}^{\ast },l)$ is
calculated within the action integral
\begin{equation}  \label{penetrability}
P(Q_{\alpha },E_{I}^{\ast },l)=\exp \left[ -\frac{2}{\hbar }\int_{R_{%
\text{in}}}^{R_{\text{out}}}\sqrt{2B(r)(E(r)-E_{sph})}dr\right]
\end{equation}
The deformation energy (relative to the sphere) is small until the rupture point
between the fragments and the two following approximations may be used: $%
R_{\text{in}}=R_{\text{d}}+R_{\alpha}$ and $B(r)=\mu$, where $\mu$ is the
reduced mass. $R_{\text{out}}$ is simply e$^{2}$Z$_{\text{d}}$Z$_{\alpha}$/Q$%
_{\alpha}$.

The residual daughter nucleus after disintegration has the most
probability to stay in its ground state, and the probability to stay
in its excited state is relatively much smaller. Therefore it is a
reasonable assumption that the probability of the residual daughter
nucleus to stay in its excited states
(\textit{I}$^{+}=2^{+},4^{+},6^{+}$,...) obeys the Boltzmann
distribution~\cite{Ren07}
\begin{equation}  \label{Boltzmann distribution}
\omega _{I}(E_{I}^{\ast })=\exp \left[ -cE_{I}^{\ast }\right]
\end{equation}
where \textit{E}$_{I}^{\ast }$ is the excitation energy of state \textit{I}$%
^{+}$ and \textit{c} is a free parameter. This excitation
probability function has been added to the model with a value of the parameter \textit{c} 
fixed to 3.0. This means that only a single parameter
is introduced in the whole calculation. It is stressed that the inclusion of
the excitation probability is reasonable in physics and it can lead to good
agreement between experiment and theory. Here \textit{I}$_{I^{+}}$ is defined 
as the product of the penetration factor and the excitation probability
\begin{equation}  \label{excitation probability}
I_{I^{+}}=\omega _{I}(E_{I}^{\ast })P(Q_{\alpha },E_{I}^{\ast },l).
\end{equation}
It is the probability of $\alpha $-transition from the
ground state of the parent nucleus to the excited states
\textit{I}$^{+}$ of the daughter nucleus. It is very convenient to
estimate the influences of these factors on the hindered $\alpha
$-transitions from \textit{I}$_{I^{+}}$. With the help of
\textit{I}$_{I^{+}}$, the branching ratios of $\alpha $-decay to
each state of the rotational band of the daughter nucleus can be
written as

\begin{equation*}
b_{\text{g.s.}}^{0^{+}}%
\%=I_{0^{+}}/(I_{0^{+}}+I_{2^{+}}+I_{4^{+}}+I_{6^{+}}+...)\times 100\%
\end{equation*}%
\begin{equation*}
b_{\text{g.s.}}^{2^{+}}%
\%=I_{2^{+}}/(I_{0^{+}}+I_{2^{+}}+I_{4^{+}}+I_{6^{+}}+...)\times 100\%
\end{equation*}%
\begin{eqnarray}
b_{\text{g.s.}}^{4^{+}}\%
&=&I_{4^{+}}/(I_{0^{+}}+I_{2^{+}}+I_{4^{+}}+I_{6^{+}}+...)\times 100\%
\notag \\
&&...
\end{eqnarray}

Similarly, the branching ratio of the $\alpha $-decay to the excited 0$^{+}$
state of the daughter nucleus is given by
\begin{equation}  \label{branching ratios 2}
b_{\text{e.s.}}^{0^{+}}=b_{\text{g.s}_{.}}^{0^{+}}\times \frac{\omega
_{0}(E_{0}^{\ast })}{\omega _{0}(0)}\frac{P_{\alpha }(Q_{\alpha
},E_{0}^{\ast },0)}{P_{\alpha }(Q_{\alpha },0,0)}
\end{equation}
where b$_{\text{g.s.}}^{0^{+}}\%$ is the branching ratio of $\alpha $%
-transition between the ground states. The $\alpha $-transition to the excited 0$%
^{+}$ state of the daughter nucleus does not involve the variation of the 
angular momentum \textit{l}, which is an ideal case for theoretical studies
of hindered $\alpha $-transitions.
\begin{figure}[htbp]
\begin{center}
\includegraphics[width=0.45\textwidth]{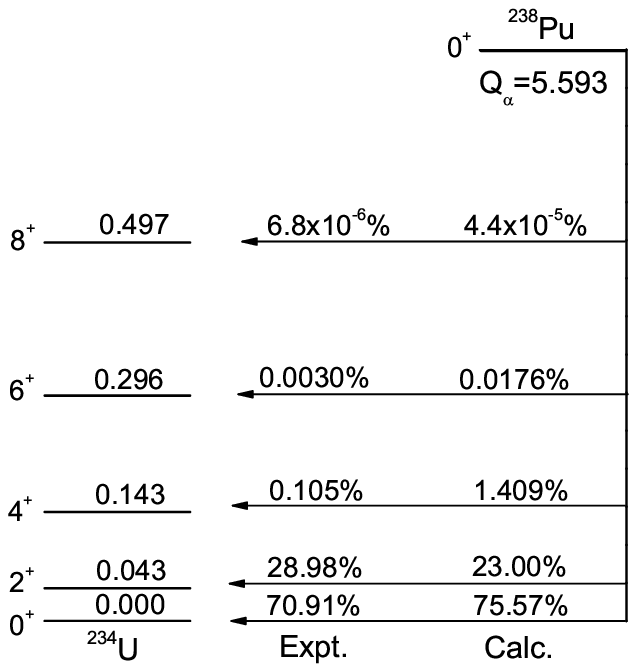}
\end{center}
\caption{The $\protect\alpha$-decay branching ratios to the
rotational band of the ground state of $^{234}$U. The $\alpha
$-decay energy \textit{Q}$_{\alpha }$ and excitation energy
\textit{E}$_{I}^{\ast }$ are measured in MeV. }
\end{figure}
\begin{figure}[htbp]
\begin{center}
\includegraphics[width=0.45\textwidth]{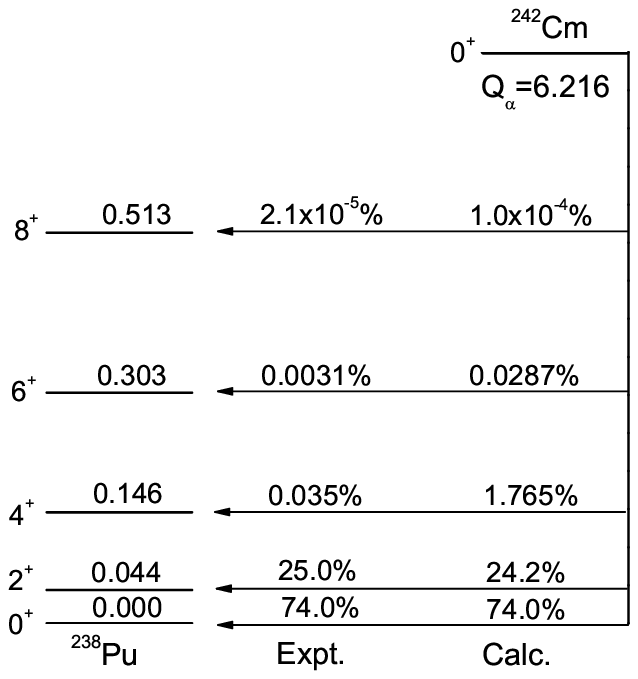}
\end{center}
\caption{The $\protect\alpha $-decay branching ratios to the
rotational band of the ground state of $^{238}$Pu. The $\alpha
$-decay energy \textit{Q}$_{\alpha }$ and excitation energy
\textit{E}$_{I}^{\ast }$ are measured in MeV. }
\end{figure}
\section{ Results and discussions}
The systematic
calculation on $\alpha $-decay branching ratios to the rotational
band is rare because some data of the excited states have been
obtained recently~\cite{Fir96}. Experimentally it is known that
the ground state of the even-even actinides mainly decays to the
0$^{+}$ and 2$^{+}$ states of
their daughter nucleus~\cite{Fir96}. The sum of branching ratios to the 0$%
^{+}$ and 2$^{+}$ states is as large as 99\% in many cases. The $\alpha $%
-transitions to other members of the rotational band (\textit{I}$%
^{+}=4^{+},6^{+},8^{+}$,...) are strongly hindered. This is different from
the $\alpha $-transition to the ground or excited 0$^{+}$ state where the
angular momentum carried by the $\alpha $-particle is zero ($\Delta l=0$).
Here the influence of the non-zero angular momentum should be included for $%
\Delta l\neq 0$ transitions. In Figs. 1-3, three typical
figures for the 
$\alpha $-decay fine structure for $^{238}$Pu, $^{242}$Cm and $^{246}$Cf are shown. The $%
\alpha $-decay branching ratios of $^{238}$Pu and $^{242}$Cm have been
measured up to 8$^{+}$ state of the rotational band, and the branching ratio
of $^{246}$Cf has been up to 6$^{+}$ state experimentally. It is shown in
these figures that the calculated values agree with the experimental ones
for both the low-lying states (0$^{+}$, 2$^{+}$) and the high-lying ones (6$%
^{+}$, 8$^{+}$), however, the calculated branching ratio to the 4$^{+}$ state is
slightly larger than the experimental one. The branching ratios to the ground state
rotational band for other even-mass $\alpha $-emitters, such as $^{224-230}$%
Th, $^{228-238}$U, $^{236-244}$Th and so on have also been calculated. The discrepancy in describing the 4$%
^{+}$ state also exists for these nuclei in the calculations. There should be unknown physics reason behind it. 
The abnormity of the 4$^{+}$ state does not allow to explain the small discrepancy. It is very interesting to pursue this by performing more
microscopic and reasonable calculations in the future. Nevertheless, the
overall agreement of branching ratios to the rotational band of
these nuclei is acceptable.

\begin{table*}[t]
\begin{ruledtabular}\caption{Experimental and calculated branching ratios of $\alpha $-decay to the excited 0$^{+}$ states of the daughter nucleus.
\textit{Q}$_{\alpha }$ is the ground state $\alpha $-decay energy
and \textit{b}$_{\text{g.s.}}^{0+}\%$ is the branching ratio of the
$\alpha $ decay to the ground state of the daughter nucleus.
\textit{E}$_{0+}^{\ast }$ is the excitation energy of the excited
0$^{+}$ state. \textit{b}$_{\text{e.s.}}^{0+}\%$ is the
corresponding experimental or theoretical $\alpha $ decay  branching
ratio.}
\begin{tabular}{c c c c c c c c c c c c c c c c}
Nuclei & \textit{Q}$_{\alpha }$ (MeV) & \textit{b}$_{\text{g.s.}}^{0+}\%$ (Expt.)& \textit{E}$_{0+}^{\ast }$ (MeV) & \textit{b}$_{\text{e.s.}}^{0+}\%$ (Expt.)& \textit{b}$_{\text{e.s.}}^{0+}\%$ (Calc.)\\
\hline
$^{226}$Th &  \ 6.452    & \ 75.5\%    &  \  0.914  & \ 3.4$\times 10^{-4}$\% & \ 1.46$\times 10^{-4}$\% \\
$^{228}$Th &  \ 5.520    & \ 71.1\%    &  \  0.916  & \ 1.8$\times 10^{-5}$\%  & \ 5.89$\times 10^{-6}$\%  \\
$^{230}$Th &  \ 4.770    & \ 76.3\%    &  \  0.825  & \ 3.4$\times 10^{-6}$\%  & \ 1.33$\times 10^{-6}$\%  \\
$^{232}$Th &  \ 4.083    & \ 77.9\%    &  \  0.721  & \ $^{\text{a}}$  & \ 3.29$\times 10^{-7}$\%    \\
$^{230}$U  &  \ 5.993    & \ 67.4\%    &  \  0.805  & \ 3.0$\times 10^{-4}$\%  & \ 1.75$\times 10^{-4}$\%  \\
$^{232}$U  &  \ 5.414    & \ 68.0\%    &  \  0.832  & \ 2.2$\times 10^{-5}$\%  & \ 1.47$\times 10^{-5}$\%  \\
$^{234}$U  &  \ 4.859    & \ 71.4\%    &  \  0.635  & \ 2.6$\times 10^{-5}$\%  & \ 1.25$\times 10^{-4}$\%  \\
$^{236}$U  &  \ 4.572    & \ 73.8\%    &  \  0.730  & \ $^{\text{a}}$  & \  3.45$\times 10^{-6}$\%  \\
$^{236}$Pu &  \ 5.867    & \ 69.3\%    &  \  0.691  & \ 6.0$\times 10^{-3}$\%  & \ 7.66$\times 10^{-4}$\%  \\
$^{238}$Pu$^{1}$ &  \ 5.593    & \ 70.9\%    &  \  0.810  & \ 5.0$\times 10^{-5}$\%  & \ 3.65$\times 10^{-5}$\%  \\
$^{238}$Pu$^{2}$ &  \ 5.593    & \ 70.9\%    &  \  1.045  & \ 1.2$\times 10^{-6}$\%  & \ 2.89$\times 10^{-7}$\% \\
$^{240}$Pu &  \ 5.256    & \ 72.8\%    &  \  0.919  & \ 6.3$\times 10^{-7}$\%  & \ 8.66$\times 10^{-7}$\%  \\
$^{242}$Pu &  \ 4.983    & \ 77.5\%    &  \  0.926  & \ $^{\text{a}}$ & \  1.78$\times 10^{-7}$\%   \\
$^{242}$Cm$^{1}$ & \ 6.216    & \ 74.0\%    &  \  0.942  & \ 5.2$\times 10^{-5}$\% & \ 1.36$\times 10^{-5}$\%  \\
$^{242}$Cm$^{2}$ & \ 6.216    & \ 74.0\%    &  \  1.229  & \ 5.1$\times 10^{-7}$\% & \ 1.02$\times 10^{-7}$\% \\
$^{244}$Cm$^{1}$ & \ 5.902    & \ 76.4\%    &  \  0.861  & \ 1.55$\times 10^{-4}$\% & \ 3.20$\times 10^{-4}$\% \\
$^{244}$Cm$^{2}$ & \ 5.902    & \ 76.4\%   &  \   1.089  & \ $^{\text{a}}$ & \  3.78$\times 10^{-7}$\% \\
\end{tabular}
\end{ruledtabular}
\end{table*}

\begin{table*}[t]
\begin{ruledtabular}\caption{The same as Table 1, but for even isotopes of Rn, Po, Pb and Hg.
The experimental data of $^{180}$Hg-$^{202}$Rn are taken from
Ref.[17]. The experimental data of $^{190}$Po are taken from Ref.
[18]. }
\begin{tabular}{c c c c c c c c c c c c c c c c}
Nuclei & \textit{Q}$_{\alpha }$ (MeV) & \textit{b}$_{\text{g.s.}}^{0+}\%$ (Expt.) & \textit{E}$_{0+}^{\ast }$ (MeV) & \textit{b}$_{\text{e.s.}}^{0+}\%$ (Expt.)& \textit{b}$_{\text{e.s.}}^{0+}\%$ (Calc.)\\
\hline
$^{180}$Hg  &  \ 6.257    & \ 33\%    &  \  0.443  & \ 2.6$\times 10^{-2}$\%  & \ 1.1$\times 10^{-1}$\% \\
$^{182}$Hg  &  \ 5.997    & \ 8.6\%    &  \  0.422  & \ 2.9$\times 10^{-2}$\% & \  2.8$\times 10^{-2}$\% \\
$^{184}$Hg  &  \ 5.658    & \ 1.25\%    &  \  0.478  & \ 2.0$\times 10^{-3}$\%  & \ 1.1$\times 10^{-3}$\% \\
$^{186}$Pb  &  \ 6.474    & \ $<$100\%    &  \  0.328  & \ $<$0.20\%   & \  1.67\% $^{\text{b}}$ \\
$^{188}$Pb  &  \ 6.110    & \ (3-10)\%    &  \  0.375  & \ (2.9-9.5)$\times 10^{-2}$\%  & \ (1.9-6.4)$\times 10^{-2}$\%   \\
$^{190}$Po$^{1}$ &  \ 7.695    & \ 96.4\%    &  \  0.523  & \ 3.3\%  & \ 0.35\%  \\
$^{190}$Po$^{2}$ &  \ 7.695    & \ 96.4\%    &  \  0.650  & \ 0.3\%  & \ 0.093\% \\
$^{194}$Po &  \ 6.986    & \ 93\%    &  \  0.658  & \ 0.22\% & \ 0.036\%  \\
$^{196}$Po &  \ 6.657    & \ 94\%    &  \  0.769  & \ 2.1$\times 10^{-2}$\%  & \ 4.8$\times 10^{-3}$\%  \\
$^{198}$Po &  \ 6.307    & \ 57\%    &  \  0.931  & \ 7.6$\times 10^{-4}$\% & \ 1.2$\times 10^{-4}$\%  \\
$^{202}$Rn &  \ 6.775    & \ (80-100)\%    &  \  0.816  & \ (1.4-1.8)$\times 10^{-3}$\% & \ (2.2-2.8)$\times 10^{-3}$\% \\
\end{tabular}
\end{ruledtabular}
\end{table*}

\begin{figure}[htbp]
\begin{center}
\includegraphics[width=0.45\textwidth]{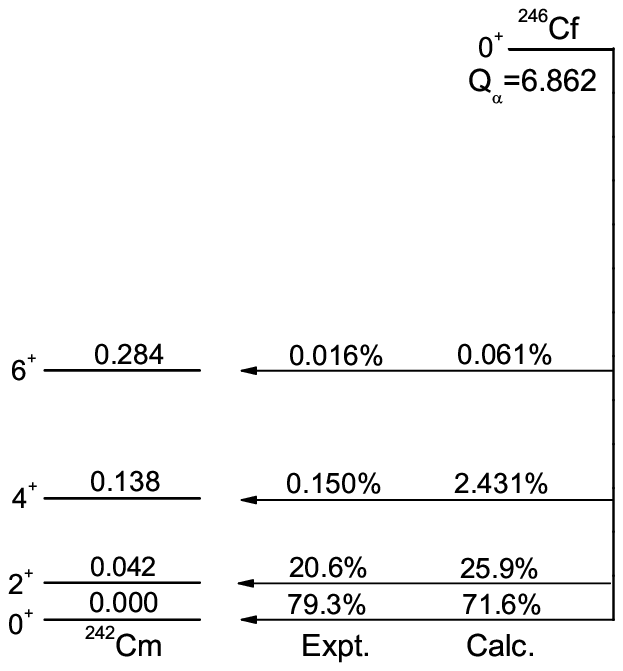}
\end{center}
\caption{The $\protect\alpha $-decay branching ratios to the
rotational band of the ground state of $^{242}$Cm. The $\alpha
$-decay energy \textit{Q}$_{\alpha }$ and excitation energy
\textit{E}$_{I}^{\ast }$ are measured in MeV.}
\end{figure}
Besides the calculations on the $\alpha $-decay branching ratios to the
rotational band, a systematic calculation on the
unfavored $\alpha $-decays to the excited 0$^{+}$ states of
even-even $\alpha $-emitters in the actinide
region has been done. The table 1 gives the experimental and calculated branching ratios of the $%
\alpha $-transition to the excited 0$^{+}$ states for even-mass
isotopes of Th, U, Pu and Cm ( in table 1, the symbol $^{\text{a}}$
represents the cases where the experimental branching ratio is
unknown ). The experimental ground state branching ratio
(b$_{\text{g.s.}}^{0^{+}}\%$) is used and its variation range
mainly from 67.4\% to 77.9\% for different nuclei in this region
~\cite{Fir96}. However, the variation of the experimental branching
ratio to the excited 0$^{+}$ state is relatively much larger and its
amplitude is as high as $0.006\%/(5.1\times 10^{-7}\%)\approx$
$10^{4}$ times (see Table 1). Therefore it is a challenging task to
obtain a quantitative agreement between experimental and theory.
The last two columns of Table 1 allow to observe that the calculated results from the GLDM are in
correct agreement with the experimental data. 
Besides the first excited 0$^{+}$ state, the $\alpha $-transitions to the second excited
0$^{+}$ state have also been observed in experiment for some nuclei,
such as $^{238}$Pu and $^{242}$Cm. Our calculated branching ratios
also agree with the experimental ones in these cases. In Table 1,
the experimental branching ratios to the first excited 0$^{+}$ state
have not been measured yet for nuclei $^{232}$Th, $^{236}$U and
$^{242}$Pu~\cite{Fir96}. The corresponding predicted values
for these nuclei are listed in Table 1. Meanwhile, the branching ratio to the
second excited 0$^{+}$ state in decay of $^{244}$Cm is also given in
Table 1. It will be very interesting to compare these theoretical
predictions with future experimental observations.

The experimental and the calculated branching ratios to the excited
0$^{+}$ states for even-mass isotopes of Hg, Pb, Po and Rn are
listed in Table 2. The hindered transitions ($\Delta l=0$) of these
nuclei involve complex particle-hole excitations above or below the
closed shell Z=82~\cite{Wod92}. Although the situation becomes more
complicated, it is seen from Table 2 that the experimental results
are reasonably reproduced in the framework of
the GLDM by taking into account the excited energies of the daughter nucleus. The $%
\alpha $-decay energies, the ground state branching ratios and the
excitation energies of nuclei in Table 2 are taken from the
experimental values~\cite{Wau94,And00}. It is found that the
calculated branching ratios of $^{202}$Rn, $^{190, 194-198}$Po,
$^{188}$Pb and $^{180-184}$Hg are consistent with the experimental
data. For the $^{186}$Pb, the calculated value
slightly deviates from the experimental one (in table 2, the symbol $^{\text{%
b}}$ denotes the cases where the calculated branching ratio deviates
from the experimental data). The agreement may be further improved
by taking into account more factors, such as the nuclear
deformations~\cite{Wod92} and so forth.

\section{conclusions}\label{sec4}
In summary, the $\alpha $-decay branching ratios of
even-even nuclei with mass numbers 180$<$ A $<$202 and A$\geq 224$ have been calculated 
in the framework of the GLDM by taking into account the angular
momentum of the $\alpha $-particle and the excitation probability of the
daughter nucleus. The calculated branching ratios to the rotational
band of the ground state of even-even actinides are consistent with
the experimental data except for the branching ratios to the 4$^{+}$ state.
The calculated branching ratios to the first and second excited
0$^{+}$ states of the daughter nucleus are also in agreement with
the available experimental data. Some predicted branching ratios are
calculated for the cases where the experimental values are still
unknown. These theoretical predictions are very helpful to the
unobserved hindered $\alpha $-transitions for future experiments.

\section{Acknowledgements}

\label{sec4} H.F Zhang is thankful to Prof. J. Q. Li, W. Zuo,
Zhong-yu Ma, Bao-qiu Chen and G. Audi for their helpful
discussions. The work is supported by the Natural Science
Foundation of China (grants 100775061, 10505016 and 10575119), by
the Fundamental Research Fund for Physics and Mathematics of
Lanzhou University (LZULL200805), by the CAS Knowledge Innovation
Project NO.KJCX-SYW-N02 and the Major State Basic Research
Developing Program of China (2007CB815004).


\end{document}